\documentclass[a4paper]{article}

\usepackage{eepic}
\usepackage{graphicx}
\usepackage{bbm}
\usepackage{amsmath}
\usepackage{alltt, amssymb}
\usepackage{amsfonts}

\usepackage{theorem}
\usepackage{graphicx}    
\usepackage{epsfig}
\usepackage{amsmath,amsfonts,latexsym,amssymb}
\usepackage{times}
\usepackage{mathrsfs}
\usepackage{algorithmic}
\usepackage{algorithm}
\usepackage{verbatim}
\usepackage{amscd}
\usepackage{pst-all}      
\usepackage{pst-poly}     

\def\C {\ensuremath{\mathbb{C}}}

\def\K {\ensuremath{\mathbb{K}}}

\def\N {\ensuremath{\mathbb{N}}}
\def\Q {\ensuremath{\mathbb{Q}}}

\def\Z {\ensuremath{\mathbb{Z}}}

\newtheorem{Theorem}{Theorem}
\newtheorem{Proposition}{Proposition}
\newtheorem{Definition}{Definition}

\newtheorem{Remark}{Remark}
\newtheorem{Example}{Example}

\newcommand{\lc}[1]{\mbox{{\rm lc}$(#1)$}}
\newcommand{\lm}[1]{\mbox{{\rm lm}$(#1)$}}
\newcommand{\lt}[1]{\mbox{{\rm lt}$(#1)$}}

\newcommand{\zero}[1]{\mbox{{\rm Zero}$(#1)$}}

\newcommand{\UniIsol}{{\small\tt UniIsol}}

\begin{document}

\title{Real Solution Isolation with Multiplicity of Zero-Dimensional
Triangular Systems}

\author{Zhihai Zhang, Tian Fang,\ Bican Xia\thanks{Corresponding author, email: xbc@math.pku.edu.cn} \\
       \small {LMAM \& School of Mathematical Sciences} \\
       \small {Peking University, Beijing 100871, China}
}
\date{}

\maketitle

\begin{abstract}

Existing algorithms for isolating real solutions of zero-dimensional
polynomial systems do not compute the multiplicities of the
solutions. In this paper, we define in a natural way the
multiplicity of solutions of zero-dimensional triangular polynomial
systems and prove that our definition is equivalent to the classical
definition of local (intersection) multiplicity. Then we present an
effective and complete algorithm for isolating real solutions with
multiplicities of zero-dimensional triangular polynomial systems
using our definition. The algorithm is based on interval arithmetic
and square-free factorization of polynomials with real algebraic
coefficients. The computational results on some examples from the
literature are presented.
\end{abstract}

\section{Introduction}\label{sec-1}

Real solution isolation for polynomials/zero-dimensional polynomial
systems/semi-algebraic systems is one of the central topics in
computational real algebra and computational real algebraic
geometry, which has many applications in various problems with
different backgrounds.

The so-called real root/zero/solution isolation of a
polynomial/zero-dimensional polynomial system/semi-algebraic system
with $k$ distinct real solutions is to compute $k$ disjoint
intervals/``boxes" containing the $k$ solutions, respectively. To
our knowledge, designing algorithms for real root isolation for
polynomials with rational coefficients was initiated by \cite{CA76}
in 1976, which was closely related to the implementation of CAD
algorithm \cite{co75}. Designing and implementation of such
algorithms have been deeply developed by many subsequent work
\cite{cl83,ABS94,Johnson,RZ04,sha} since then. Those algorithms are
mainly based on Descartes' rule of sign or Vincent's theorem.

To generalize the algorithms for polynomials to zero-dimensional
triangular polynomial systems, one must consider real root isolation
for polynomials with real algebraic coefficients. There are indeed
some work to generalize Descartes' rule of sign to polynomials with
algebraic coefficients. However, dealing with algebraic coefficients
directly may affect efficiency greatly.

In \cite{xy02,xz06} we considered real solution isolation for
semi-algebraic systems with finite solutions. We introduced a method
which always enables us to avoid handling directly polynomials with
algebraic coefficients and to deal with polynomials with rational
coefficients only. A recent algorithm in \cite{CGY07} can compute
the parity of the solutions as well as isolate real roots of
zero-dimensional triangular polynomial systems.

In this paper, we define in a natural way the multiplicity of
solutions of zero-dimensional triangular polynomial systems and
prove that our definition is equivalent to the classical definition
of local (intersection) multiplicity. Then we present an effective
and complete algorithm for isolating real solutions with
multiplicities of zero-dimensional triangular polynomial systems
using our definition. The algorithm is based on square-free
factorization of polynomials with real algebraic coefficients and
our previous work \cite{xz06}. We also provide computational results
on some examples from the literature.

In this paper, all polynomials are in $\C[X]=\C[x_1,\ldots,x_n]$ if
not specified.

\section{Multiplicities of zeros of triangular sets}

First, let's recall the definition of {\em local (intersection)
multiplicity}. We follow the notations in Chapter 4 of \cite{cox}.
Although some notations and definitions can be stated in a more
general way, we restrict ourselves to the ring
$\C[X]=\C[x_1,\ldots,x_n]$ since we are interested in the complex or
real zeros of zero-dimensional polynomial systems.

For $p=(\eta_1,\ldots,\eta_n)\in \C^n$, we denote by $M_p$ the
maximal ideal generated by $\{x_1-\eta_1,\ldots,x_n-\eta_n\}$ in
$\C[X]$, and write
\[\C[X]_{M_p}=\left\{\frac{f}{g}: f,g \in \C[X], g(\eta_1,\ldots,\eta_n)\ne0\right\}.\]
It is well-known that $\C[X]_{M_p}$ is the so-called {\em local
ring}.

\begin{Definition}{\rm \cite{cox}}\label{cox}
Suppose $I$ is a zero-dimensional ideal in $\C[X]$ and $p\in \zero
I$, the zero set of $I$ in $\C$. Then the {\em multiplicity} of $p$
as a point in $\zero I$ is defined to be
\[\dim_k\C[X]_{M_p}/I\C[X]_{M_p}.\]
That is, the multiplicity of $p$ is the dimension of the quotient
space $\C[X]_{M_p}/I\C[X]_{M_p}$ as a vector space over $\C$.
\end{Definition}

For a zero of a zero-dimensional triangular set, there can be a
natural and intuitional definition of multiplicity as follows.

\begin{Definition}\label{our}
For a zero-dimensional triangular system,
\begin{equation*}
\left\{\begin{array}{l}f_1(x_1)=0,\\
f_2(x_1,x_2)=0,\\
\ldots\\
f_n(x_1,\ldots,x_n)=0,
\end{array}
\right.
\end{equation*}
and one of its zeros, $\xi=(\xi_1,\ldots,\xi_n)$, the {\em
multiplicity} of $\xi$ is defined to be $\prod\limits_{i=1}^{n}m_i$,
where $m_i$ is the multiplicity of $x_i=\xi_i$ as a zero of the
univariate polynomial $f_i(\xi_1,\ldots,\xi_{i-1},x_i)$ for
$i=1,\ldots,n$.
\end{Definition}

\begin{Example}
Consider the following triangular system:
\begin{equation*}
\left\{\begin{array}{l}g_1=x_1^3+2x_1^5+7x_1^7=0,\\
g_2=x_2^3+x_2^2+x_1x_2=0,\\
g_3=x_3^2+x_1x_3+x_1x_2=0.
\end{array}
\right.
\end{equation*}
Let's compute the local multiplicity of $(0,0,0)$ by Definition
\ref{our}. The multiplicity of $x_1=0$, a zero of $g_1$, is $3$.
Substitute $x_1=0$ in $f_2$ and the resulted $g_2$ is $g_2'=x_2^2$.
Thus, the multiplicity of $x_2=0$, a zero of $g_2'$, is $2$.
Finally, substitute $x_1=x_2=0$ in $g_3$, and the resulted $g_3$ is
$g_3'=x_3^2$. Thus, the multiplicity of $x_3=0$, a zero of $g_3'$ is
$2$. As a result, the local multiplicity of $(0,0,0)$ is $3\times
2\times 2=12$.
\end{Example}

In the following, we will prove that Definition \ref{our} is
equivalent to Definition \ref{cox}. Many notations and results are
taken from \cite{cox}.

Usually, a total order that is compatible with multiplication and
that satisfies $1>x_i$ for all $i$'s, is called a {\em local order}.

\begin{Definition}\cite{gg}({\bf Negative lexicographical ordering})\label{nl}\\
Assume $\alpha=(\alpha_1,\ldots,\alpha_n)\in\N_{\ge}^n$ and
$\beta=(\beta_1,\ldots,\beta_n)\in\N_{\ge}^n$. We say
$X^\alpha>_{nl}X^\beta$ if
\[\exists i~(1\le i\le n)\wedge (\forall j,~ 1\le j<i\Longrightarrow
\alpha_j=\beta_j)\wedge(\alpha_i<\beta_i).\]
\end{Definition}

\begin{Remark}
The negative lexicographical ordering $>_{nl}$ is obviously a local
order.
\end{Remark}


For a given order, $\lc{f}, \lm{f}$ and $\lt{f}$ denote the {\em
leading coefficient}, {\em leading monomial} and {\em leading term}
of $f$, respectively. For a set $S$, $\lt{S}=\{\lt{f}: f\in S\}$.

\begin{Definition}{\rm \cite{cox}}
Let $R=\mathbb{C}[X]_{M_p}$ and $I\subset R$ be an ideal. A set
$\left\{ g_1,\ldots,g_m\right\}\subset I$ is called a {\em standard
basis} for $I$ with respect to $<_{nl}$ if $\langle \lt
I\rangle=\langle \lt{g_1},\ldots,\lt{g_m}\rangle.$
\end{Definition}

For $\alpha=(\alpha_1,\ldots,\alpha_n)\in\N_{\ge}^n$, define
$|\alpha|=\sum_{i}\alpha_i.$ For any polynomial $g=\sum_\alpha
c_\alpha X^{\alpha}\in \mathbb{C}[X]$ with total degree $d$, we will
write $g^h=\sum_\alpha c_\alpha t^{d-|\alpha|}X^{\alpha}$ for the
homogenization of $g$ with respect to $t$. 

\begin{Definition}{\rm \cite{cox}}
Define $t^aX^{\alpha}>_{nl}'t^bX^{\beta}$ if $a+|\alpha|>b+|\beta|$
or $a+|\alpha|=b+|\beta|$, but $X^\alpha>_{nl}X^{\beta}$.
\end{Definition}

It is easy to verify that $>_{nl}'$ is a monomial order over
$\C[t,X]$.

\begin{Theorem}{\rm \cite{cox}}\label{thm:cox}(Analog of Buchberger's Criterion)\\
Let $G=\left\{g_1,\ldots,g_m\right\}$, $>$ be any local order, and
$I$ be the ideal in $\mathbb{C}[X]_{M_p}$ generated by $G$. $G$ is a
standard basis for $I$ if and only if applying Mora normal form
algorithm to each S-polynomial formed from elements of the set of
homogenizations $G^h=\left\{g_1^h,\ldots,g_m^h\right\}$ yields a
zero remainder.
\end{Theorem}


For our purpose, we state the criterion in another form as follows.

\begin{Theorem}\label{thm:standard basis}
Let notations be as in Theorem \ref{thm:cox}. $G$ is a standard
basis if and only if for any nonzero S-polynomial of $g_i^h$ and
$g_j^h$, denoted by $S_{ij}$, there exist homogeneous polynomials
$U, A_1,\ldots, A_m\in \mathbb{C}[t,X]$ such that
\begin{equation}\label{sij}US_{ij}=\sum\limits_{l=1}^{m}A_lg_l^h,\end{equation}
where $\lt{U}=t^a$ for some $a$,
\[a+\deg(S_{ij})=\deg(A_l)+\deg(g_l^h)\]
for all $l$ whenever $A_l\ne0$, and
$\lt{A_lg_l^h}\leq_{nl}'\lt{US_{ij}}$.
\end{Theorem}

\begin{Remark}
We omit the proof of Theorem \ref{thm:standard basis}, which is
almost the same as that of Theorem \ref{thm:cox}. The criterion in
Theorem \ref{thm:standard basis} is independent to any algorithms.
One can use Mora normal form algorithm to get such representation as
(\ref{sij}) for each $S_{ij}$ if $G$ is a standard basis. 

\end{Remark}

Without loss of generality, in the rest of this section we assume
$p=(0,\ldots,0)$ is a zero of the triangular set under discussion
and focus on its multiplicity. Consider the following triangular set
with leading terms $c_1x_1^{m_1},\ldots,c_nx_n^{m_n}$ respectively
w.r.t. the order $>_{nl}$:
\begin{equation}\label{T}T=
\left\{\begin{array}{l}f_1(x_1)=c_1x_1^{m_1}+t_1(x_1),\\
f_2(x_1,x_2)=c_2x_2^{m_2}+t_2(x_1,x_2),\\
\ldots,\\
f_n(x_1,\ldots,x_n)=c_nx_n^{m_n}+t_n(x_1,\ldots,x_n),
\end{array}
\right.
\end{equation}
where $t_i(x_1,\ldots,x_i)$ is a polynomial in $x_1,\ldots,x_i$ for
$i=1,\ldots,n$ and $c_i$'s are constants. 
Without loss of generality, we assume $c_i$'s are all $1$ in the
proof of the following proposition.

\begin{Proposition}\label{pro:standard basis} Let $T$ be as above and $I=\langle
T\rangle$ the ideal generated by $T$ in the local ring
$\C[X]_{\langle x_1,\ldots,x_n\rangle}$. Then $T$ is a standard
basis for $I$ with respect to $>_{nl}$.
\end{Proposition}
\begin{Proof}
According to Theorem \ref{thm:standard basis}, we only need to show
that every nonzero S-polynomial of each pair of
$T^h=\left\{f_1^h,\ldots,f_n^h\right\}$ can be represented in the
form of (\ref{sij}).

Assume that $f_i^h=t^a x_i^{m_i}+\overline{t_i}, f_j^h=t^b
x_j^{m_j}+\overline{t_j}$ and $a<b$. The S-polynomial, $S_{ij}$, of
$f_i^h$ and $f_j^h$ is
\[S_{ij}=t^{b-a}x_j^{m_j}f_i^h-x_i^{m_i}f_j^h.\]
Let $p_1=t^{b-a}x_j^{m_j}f_i^h$ and $p_2=x_i^{m_i}f_j^h$. Under the
order $<'_{nl}$, the first term of $p_1$ is equal to the first term
of $p_2$. If $S_{ij}\ne0$, there exists some $L$ such that under the
order $<'_{nl}$ the $L$-th term of $p_1$ is not equal to the $L$-th
term of $p_2$ and the $k$-th term of $p_1$ is equal to the $k$-th
term of $p_2$ for all $1\le k<L$. Then, the $k$-th terms of $p_1$
and $p_2$ can be represented as $x_i^{m_i}q_k$ and
$t^{b-a}x_j^{m_j}q_k$ for some $q_k$, respectively. Thus, $f_i^h$
and $f_j^h$ can be respectively rewritten as
\[f_i^h=x_i^{m_i}(t^a+Q)+\overline{f_{i2}},
f_j^h=t^{b-a}x_j^{m_j}(t^{a}+Q)+\overline{f_{j2}},\] where
$Q=\sum\limits_{k=1}^{L-1}q_k$ and $\overline{f_{i2}}$ and
$\overline{f_{j2}}$ are the remained parts of $f_i^h$ and $f_j^h$,
respectively, which satisfy that
\[\lt{t^{b-a}x_j^{m_j}\overline{f_{i2}}}\ne
\lt{x_i^{m_i}\overline{f_{j2}}}.\] It is easy to verify that
$S_{ij}=t^{b-a}x_j^{m_j}\overline{f_{i2}}-x_i^{m_i}\overline{f_{j2}}.$
Then
\[\lm{S_{ij}}=\max(\lm{t^{b-a}x_j^{m_j}\overline{f_{i2}}},\lm{x_i^{m_i}\overline{f_{j2}}})\]
under the order $<'_{nl}$. Thus,
\[(t^a+Q)S_{ij}=\overline{f_{i2}}f_j^h-\overline{f_{j2}}f_i^h.\]
Let $U=t^a+Q, A_j=\overline{f_{i2}}, A_i=\overline{f_{j2}}$ and
$A_{l}=0 (l\ne i~ {\rm and}~ l\ne j)$. Then we have
\[US_{ij}=\sum\limits_{l=1}^nA_lf_l^h,\]
and all the requirements in Theorem \ref{thm:standard basis} are
met. Thus $T$ is a standard basis for $\langle T\rangle$ w.r.t
$<_{nl}$.
\end{Proof}

In order to prove the equivalence of Definitions \ref{our} and
\ref{cox} about the (local) multiplicity, we need the following
theorem, which can be found in \cite{cox}.

\begin{Theorem}{\rm \cite{cox}}\label{book}
Let $I$ be an ideal in a local ring $R$, and assume that $\dim_k
~R/\langle \lt{I}\rangle$ is finite for some local order $>$. Then
we have
\[\dim_k~R/I=\dim_k~R/\langle \lt{I}\rangle.\]
\end{Theorem}

\begin{Theorem}\label{equivalence}
Let notations be as above and $T$ a zero-dimensional triangular set
with a zero $p=(0,\ldots,0)$. If the multiplicity of $p$ defined by
Definition \ref{our} is $m=\prod\limits_{i=1}^{n}m_i$, then the
local multiplicity, defined by Definition \ref{cox}, of $p$ as a
point of $\langle T\rangle$ is also $m$.
\end{Theorem}
\begin{Proof} If the multiplicity is $\prod\limits_{i=1}^{n}m_i$ in the sense of
Definition \ref{our}, then $T$ can be rewritten as
\begin{equation*}T=
\left\{\begin{array}{l}f_1(x_1)=(c_1+t_{11}(x_1))x_1^{m_1},\\
f_2(x_1,x_2)=(c_2+t_{21}(x_1,x_2))x_2^{m_2}+x_1t_{22}(x_1,x_2),\\
\ldots,\\
f_n(X)=(c_n+t_{n1}(X))x_n^{m_n}+\sum\limits_{i=1}^{n-1}x_it_{ni+1}(X),
\end{array}
\right.
\end{equation*}
where $X=(x_1,\ldots,x_n)$, the $t_{ij}(X)$s are polynomials in
$(x_1,\ldots,x_i)$ and the $t_{i1}(X)$s do not contain constants.

Under the order $>_{nl}$, the leading term of $f_i(x_1,\ldots,x_i)$
is $c_ix_i^{m_i}$ for $i=1,\ldots,n$. According to Proposition
\ref{pro:standard basis}, $T$ is a standard basis of $I=\langle
T\rangle.$ Thus,
\[\langle \lt{I}\rangle=\langle x_1^{m_1},\ldots,x_n^{m_n}\rangle.\]

Let $R=\C[X]_{\langle x_1,\ldots,x_n\rangle}.$ According to Theorem
\ref{book},
\[\dim_kR/IR=\dim_kR/\langle
x_1^{m_1},\ldots,x_n^{m_n}\rangle R=\prod\limits_{i=1}^{n}m_i.\]
\end{Proof}



\section{Algorithm for real solution isolation with multiplicity}

In this section, based on the results in last section, we present an
algorithm for real root isolation with multiplicity of
zero-dimensional triangular polynomial equations. That is, we not
only isolate the real roots, but also compute the multiplicity of
each real root by Definition \ref{our} at the same time. In this
section, the input polynomial or polynomial set to our algorithms is
taken from $\Q[X]$.

It is well-known that there exist some efficient algorithms for real
root isolation of polynomials or polynomial equations or
semi-algebraic systems
\cite{CA76,cl83,ABS94,Johnson,xy02,RZ04,CGY07,xz06}. To obtain the
multiplicities of the real roots at the same time, our idea is
simple that is to take use of {\em square-free factorization} of
polynomials with rational or algebraic coefficients. When dealing
with algebraic coefficients, we make use of the idea in
\cite{xy02,xz06} which enables us to deal with rational coefficients
instead.

For the univariate case, suppose $p=\prod_{i=1}^{k}p_i^i$. Isolating
the real zeros of $p$ with multiplicity contains two main steps. One
is to compute the squarefree factorization of $p$, the other is to
isolate the real zeros of the squarefree part of $p$. We can use
many existing tools to obtain the squarefree factorization, {\it
i.e.}, those $p_i$s. Then we know at once the multiplicities of
those real zeros of each $p_i$. In principle, we may isolate the
real zeros of the squarefree part of $p$ in two ways. One way is to
isolate the real zeros of $p_1p_2\cdots p_k$ first and then match
the zeros with $p_i$ to obtain correct multiplicities. The other way
is to isolate the real zeros of each $p_i$ separately. However, in
the later way, we may need to compute a root gap of $p$ first.
Anyway, the univariate case can be efficiently dealt with. So, we do
not enter the details of such algorithms and only give a description
of the input and output of such function.


\medskip
\noindent\textbf{Calling sequence} ~~ ${\UniIsol}(f(x))$\\
\textbf{Input:} a univariate polynomial $f(x)$\\
\textbf{Output:} a set of elements of the form $([a,b],m)$ where
$[a,b]$ is an interval containing exact one real root of $f(x)=0$
and $m$ is the multiplicity of the root. There are not any real
roots of $f(x)=0$ outside the intervals.
\medskip

Then let us consider the multivariate case. To be more precise, we
state our problem as follows: that is to isolate the real solutions
with multiplicities of the following zero-dimensional triangular
polynomial set
\[T=\{f_1(x_1),f_2(x_1,x_2),\ldots,f_n(x_1,\ldots,x_n)\}.\]
In principle, Definition \ref{our} suggests a naive method to
compute the local multiplicity as follows. First compute all the
zeros of $f_1(x_1)$ and their multiplicities by $\UniIsol$; then
``substitute" the zeros for $x_1$ in $f_2(x_1,x_2)$ one by one, and
compute all the zeros of the resulted $f_2(\bar{x_1},x_2)$ and their
multiplicities by ${\UniIsol}$ again, and so on. Of course, in
general we cannot directly substitute the zeros in those polynomials
because they may be algebraic numbers of high degrees. Nevertheless,
this naive method is the main framework of our algorithm.


Let $T_i=\{f_1(x_1),f_2(x_1,x_2),\ldots,f_i(x_1,\ldots,x_i)\}$. We
will call
\[([a_1,b_1],\ldots,[a_i,b_i])~~{\rm or}~~([a_1,b_1],\ldots,[a_i,b_i],m)\] an
{\em interval solution} of $T_i$ (with multiplicity $m$) if the
``box" $[a_1,b_1]\times\cdots\times[a_i,b_i]$ contains exact one
real solution of $T_i$ (and $m$ is the multiplicity of the
solution). If $T_i$ has $k$ distinct real solutions, a set of $k$
interval solutions of $T_i$ containing respectively the $k$ real
solutions is called a {\em solution set} of $T_i$. For an interval
solution $r=([a_1,b_1],\ldots,[a_i,b_i])$, we define $N_r=[x_1-a_1,
b_1-x_1, \ldots, x_i-a_i, b_i-x_i]$ and $N_r\ge 0$ stands for
$a_1\le x_1\le b_1,\ldots, a_i\le x_i\le b_i$, i.e., $(x_1,\ldots,
x_i)\in [a_1,b_1]\times\cdots\times[a_i,b_i]$.

Suppose we already have a solution set of $T_i$ and
\[(\xi_1,\ldots,\xi_i)\in [a_1,b_1]\times\cdots\times[a_i,b_i]\]
is a real root of $T_i$ with multiplicity $m$. To isolate the real
zeros of $f_{i+1}(\xi_1,\ldots,\xi_i,x_{i+1})$ with multiplicity, we
need to \begin{enumerate}
\item compute the algebraic squarefree factorization of
$f_{i+1}(\xi_1,\ldots,\xi_i,x_{i+1})$, and
\item isolate the real zeros of the squarefree part computed.
\end{enumerate}

Let us first consider the second task, i.e., how to isolate the real
zeros of $f_{i+1}(\xi_1,\ldots,\allowbreak \xi_i,x_{i+1})$ if it is
squarefree. In \cite{xz06}, we proposed a complete algorithm, called
${\small\tt RealZeros}$, for isolating the real solutions (without
multiplicities) of semi-algebraic systems. Our second task can be
accomplished by a sub-algorithm of ${\small\tt RealZeros}$. The key
idea of the algorithm is to compute two suitable polynomials
$\overline{f_{i+1}}$ and $\underline{f_{i+1}}$ in $x_{i+1}$ with
rational coefficients such that
\[\underline{f_{i+1}}<f_{i+1}(\xi_1,\ldots,\xi_i,x_{i+1})<\overline{f_{i+1}}\]
by using interval arithmetic and those intervals
$[a_1,b_1],\ldots,[a_i,b_i].$ And the real zeros of
$f_{i+1}(\xi_1,\ldots,\xi_i,x_{i+1})$ can be isolated through
isolating the real zeros of $\overline{f_{i+1}}$ and
$\underline{f_{i+1}}$. Therefore, we can avoid dealing with
polynomials with algebraic coefficients directly. In the following,
we call this sub-algorithm ${\small\tt AlgebraicIsolate}$.

\medskip
\noindent\textbf{Calling sequence} ~~
${\small\tt AlgebraicIsolate}(g(x_1,\ldots,x_{i+1}),T_{i},r)$\\
\textbf{Input:} a squarefree polynomial $g(x_1,\ldots,x_{i+1})$, a
zero-dimensional triangular polynomial set $T_{i}$ as above and an
interval solution $r=([a_1,b_1],\ldots,[a_i,b_i])$ which contains
exact one real
zero $(\xi_1,\ldots,\xi_i)$ of $T_{i}$.\\
\textbf{Output:} a list of isolating intervals of real zeros of
$g(\xi_1,\ldots,\xi_i,x_{i+1})$.
\medskip

For the detail of the algorithm ${\small\tt AlgebraicIsolate}$,
please be referred to \cite{xz06}.

Now, we turn to the first task, i.e., compute the algebraic
squarefree factorization of $f_{i+1}(\xi_1,\ldots,\xi_i,x_{i+1})$.
One may use some existing algorithms for algebraic factorization,
see for example \cite{wang}, to accomplish the task. In the
following, we propose a method for algebraic squarefree
factorization based on algebraic gcd computation. A key manipulation
in the computation is to count real solutions of semi-algebraic
systems by an algorithm ${\small\tt RealrootCounting}$ in
\cite{xh02}\footnote{The algorithm is called ${\tt nearsolve}$ in
\cite{xh02}}.

\medskip
\noindent\textbf{Calling sequence} ~~ ${\small\tt RealrootCounting}(F, N, P, H)$\\
\textbf{Input:} a zero-dimensional polynomial set $F$, a list of
non-strict inequalities $N$, a list of strict inequalities $P$ and a
list of inequations $H$.\\
\textbf{Output:} the number of real roots of the system $\{F=0, N\ge
0, P>0, H\ne 0\}$.
\medskip


\medskip
\noindent\textbf{Calling sequence} ~~
${\small\tt AlgebraicGCD}(p_1(x_1,\ldots,x_{i+1}),p_2(x_1,\ldots,x_{i+1}),T_i,r)$\\
\textbf{Input:} two polynomials $p_1,p_2$ in $x_1,\ldots,x_{i+1}$, a
zero-dimensional triangular polynomial set $T_i$ as above and an
interval
solution $r=([a_1,b_1],\ldots,[a_i,b_i])$ of $T_i$.\\
\textbf{Output:} The greatest common divisor of $p_1$ and $p_2$
viewed as polynomials in $x_{i+1}$ w.r.t. the interval solution $r$,
i.e.,
$\gcd(p_1(\xi_1,\ldots,\xi_i,x_{i+1}),p_2(\xi_1,\ldots,\xi_i,x_{i+1}))$
where $(\xi_1,\ldots,\xi_i)$ is the only real solution in $r$.

\begin{description}
\item[Step 0] Suppose the subresultant chain of $p_1$ and
$p_2$ w.r.t. $x_{i+1}$ is $S_{\mu}, S_{\mu-1},\ldots, S_0$ with
principal subresultant coefficients $R_{\mu}, R_{\mu-1},\ldots,
R_0$, respectively. Set $j\leftarrow 0$.

\item[Step 1]  Compute $R_j$.

\item[Step 2] If ${\small\tt RealrootCounting}(T_i,N_r,[~],[R_j])=0$, i.e., the
interval solution makes $R_j$ vanish, then set $j\leftarrow j+1$ and
go to Step 1.

\item[Step 3] Return $S_j$.
\end{description}

There are several mature algorithms \cite{gcl92,vg99} for squarefree
factorization of polynomials in $\K[x]$ where $\K$ is $\Z, \Q$ or a
finite field. It is well known that such algorithms for univariate
case only contain two main manipulation: gcd computation and
polynomial division in $\K[x].$ If we replace the gcd computation in
those algorithms with our ${\small\tt AlgebraicGCD}$ computation and
replace the division manipulation with pseudo-division, then those
algorithms will compute algebraic squarefree factorization as we
want. So, We only give a simple description of our algorithm here.

\medskip
\noindent\textbf{Calling sequence} ~~
${\small\tt AlgebraicSqfreeFactor}(p(x_1,\ldots,x_{i+1}),T_i,r)$\\
\textbf{Input:} a polynomial $p$ in $x_1,\ldots,x_{i+1}$, a
zero-dimensional triangular polynomial set $T_i$ as above and an
interval solution $r=([a_1,b_1],\ldots,[a_i,b_i])$ of $T_i$.\\
\textbf{Output:} the squarefree factorization of $p$ viewed as a
polynomial in $x_{i+1}$ w.r.t. the interval solution $r$, i.e., the
squarefree factorization of $p(\xi_1,\ldots,\xi_i,x_{i+1})$ where
$(\xi_1,\ldots,\xi_i)$ is the only real solution in $r$.
\medskip

Now, we are ready to describe our algorithm ${\small\tt
MultiIsolate}$ for real solution isolation with multiplicity of
zero-dimensional triangular polynomial sets.

\medskip
\noindent\textbf{Calling sequence} ~~
${\small\tt MultiIsolate}(T)$\\
\textbf{Input:} a zero-dimensional triangular polynomial set
$T=\{f_1(x_1),\ldots,f_n(x_1,\ldots,x_n)\}$.\\
\textbf{Output:} a solution set of $T$ with multiplicity.

\begin{description}
\item[Step 1] $i\leftarrow 1$, $L_i\leftarrow \UniIsol(f_1)$.

\item[Step 2] $L_i$ is a solution set of $T_i$ with multiplicity.
If $i=n$, return $L_n$.

\item[Step 3] For each interval solution $r=([a_1,b_1],\ldots,[a_i,b_i])$
in $L_i$ with multiplicity, compute ${\small\tt
AlgebraicSqfreeFactor}(f_{i+1}(x_1,\ldots,x_{i+1}),T_i,r)$. So, we
know at once the multiplicity of each factor. Assume
$\widetilde{f_{i+1}}$ is the squarefree part of $f_{i+1}$. Then, by
applying ${\small\tt AlgebraicIsolate}(\widetilde{f_{i+1}}, T_i, r)$
we can obtain the isolating intervals of real zeros of $f_{i+1}$.
So, it is easy to obtain a solution set $L_{i+1}$ of $T_{i+1}$ with
multiplicity by Definition \ref{our}.

$i\leftarrow i+1$ and go to Step 2.

\end{description}

\begin{Remark}\label{mo}
Let $r=([a_1,b_1],\ldots,[a_n,b_n])$ be an interval solution of $T$
and $\xi=(\xi_1,\ldots,\xi_n)$ is the real solution in $r$. If
$\lc{f_i}(\xi_1,\ldots,\xi_{i-1})\ne 0$ for $2\le i\le n$, $T$ is
said to be {\em regular} w.r.t. $\xi$ (or $r$). If
$f_i(\xi_1,\ldots,\xi_{i-1},x_i)$ is squarefree for $1\le i\le n$,
$T$ is said to be {\em squarefree} w.r.t. $\xi$ (or $r$).

It is clear that ${\small\tt MultiIsolate}(T)$ actually computes as
well a regular and squarefree decomposition of the given triangular
set $T$ w.r.t. its real zeros, respectively. That is to say, we
compute a set of triangular sets $W_j$ and their solution sets $Q_j$
such that $\cup_{j}Q_j$ is a solution set of $T$ and each $W_j$ is
regular and squarefree w.r.t. each solution in $Q_j$. If we modify
slightly the algorithm, we can output the regular and squarefree
decomposition.
\end{Remark}


\section{Examples}

The algorithm ${\small\tt MultiIsolate}$ has been implemented as a
Maple program which is included in our package DISCOVERER
\cite{xia07}. For an input zero-dimensional triangular system, our
program can compute the real solution isolation of the system with
multiplicity and output a regular and squarefree decomposition (see
Remark \ref{mo}) of the system w.r.t. those real solutions. Our
program can detect whether the input system is zero-dimensional. If
it is not, the program will return a message: ``The dimension of the
system is positive."

In this section, we illustrate the function of our program by some
examples. The timings are collected on a Thinkpad X200 running Maple
11 with 1G memory and Windows Vista.


\begin{Example}
Consider the following triangular system:
\begin{equation*}
\left\{\begin{array}{l}f_1=x-2,\\
f_2=(x+y-3)^3(y+3),\\
f_3=(yz^2+xz+1)^2((x-y)^4z+x-y).
\end{array}
\right.
\end{equation*}
Within $1.6$ seconds, our program outputs a solution set as follows.
\begin{eqnarray*} && [~~[~ [[2, 2],
[-3,-3], [-\frac{1}{4}, 0]], 1~],  [~ [[2, 2], [-3, -3], [1, 1]], 2~],\\
&& [~ [[2, 2], [-3, -3], [-\frac{1}{2}, -\frac{1}{4}]], 2~],  [~
[[2, 2], [1, 1], [-1, -1]], 15~] ~~].
\end{eqnarray*}
That means the system has $4$ real solutions which are of
multiplicities $1,2,2,15,$ respectively. Our program also outputs a
regular and squarefree decomposition of the system w.r.t. the four
distinct real solutions respectively as follows.
\begin{eqnarray*}
&&[x-2, y+3, 1+125z],\\
&&[x-2, y+3, -1+3z^2-2z],\\
&&[x-2, y-1, z+1].
\end{eqnarray*}
Note that the second and third solutions are both solutions to the
second equations above.
\end{Example}

\begin{Example}
Consider the following triangular system:
\begin{equation*}
\left\{\begin{array}{l}f_1=(x+1)(x-2),\\
f_2=(x-y+1)^2(y-5)+(y-3)x,\\
f_3=(xy-6)z^2+2z+1.
\end{array}
\right.
\end{equation*}
The system has $7$ real solutions all of multiplicities $1$. The
computation costs $0.7$ seconds.
\begin{eqnarray*}
&&[~~[~ [[2, 2], [3, 3], [-1/2, -1/2]], 1~], [~[[-1, -1], [-1,
-3/4],
[-3/8, -1/8]], 1~],\\
&&[~ [[-1, -1], [-1, -3/4], [3/8, 7/8]], 1~ ], [~[[-1, -1], [1/2,
3/4],
[-3/8, -1/8]], 1~ ],\\
&&[~[[-1, -1], [1/2, 3/4], [3/8, 3/4]], 1~], [~[[-1, -1], [5, 21/4],
[-3/8, -1/8]], 1~],\\
&&[~[[-1, -1], [5, 21/4], [1/4, 1/2]], 1~]~~].
\end{eqnarray*}
A regular and squarefree decomposition is
\[[x-2, f, 1+2z],~[x+1, f, g],\]
where $f=x^2y-5x^2-2xy^2+13xy-13x+y^3-7y^2+11y-5,$
$g=yz^2x-6z^2+2z+1$.
\end{Example}

\begin{Example}
The following triangular system is taken from \cite{zhonggang}.
\begin{equation*}
\left\{\begin{array}{l}f_1=x^4,\\
f_2=x^2y+y^4,\\
f_3=z+z^2-7x^3-8x^2.
\end{array}
\right.
\end{equation*}
Within $0.1$ seconds, we obtain two distinct real roots with
multiplicities $16$.
$$[~~[~[0, 0], [0, 0], [-1, -1]~], 16~~],$$
$$[~~[~[0, 0], [0, 0], [0, 0]~], 16~~]. $$
And a regular and squarefree decomposition is
$$[x, y, z+z^2-7x^3-8x^2].$$
\end{Example}

\begin{Example}
The following triangular system is taken from \cite{CGY07}.
\begin{equation*}
\left\{\begin{array}{l}f_1=x^4-3x^2-x^3+2x+2,\\
f_2=y^4+xy^3+3y^2-6x^2y^2+4xy+2xy^2-4x^2y+4x+2.
\end{array}
\right.
\end{equation*}
The time for computation is $3.6$ seconds and we obtain $12$
distinct real roots.
$$[~~[~[51/32, 13/8], [-119/32, -475/128]~], 1~~],$$ $$
[~~[~[51/32, 13/8], [-147/128, -145/128]~], 1~~],$$ $$ [~~[~[51/32,
13/8], [53/64, 107/128]~], 1~~],$$ $$ [~~[~[51/32, 13/8], [307/128,
77/32]~], 1~~],$$ $$ [~~[~[-5/8, -19/32], [-3/8, 1/4]~], 1~~],$$ $$
[~~[~[-5/8, -19/32], [13/8, 17/8]~], 1~~],$$ $$ [~~[~[45/32, 23/16],
[-3025/1024, -1499/512]~], 1~~],$$
$$ [~~[~[45/32, 23/16], [-1347/1024, -2639/2048]~], 1~~],$$ $$
[~~[~[45/32, 23/16], [11/8, 3/2]~], 2~~],$$ $$ [~~[~[-23/16,
-45/32], [-5/8, -1/8]~], 1~~],$$ $$ [~~[~[-23/16, -45/32], [17/4,
5]~], 1~~],$$ $$ [~~[~[-23/16, -45/32], [-3/2, -11/8]~], 2~~].$$ It
is clear that two of the solutions are of multiplicities $2$ and the
others are of multiplicities $1$. With respect to those solutions,
we have a regular and squarefree decomposition as follows.
$$[x^2-x-1, h_1],~[x^2-2,h_2],~[x^2-2, h_3],$$
where
\begin{eqnarray*}
&&h_1=y^4+xy^3+3y^2-6x^2y^2+4xy+2xy^2-4x^2y+4x+2,\\
&&h_2=-23354573041809-9122537689096xy^2+39406733143725xy+\\
&&~~~~~~~~17148617740054x+13135577714575y^2-54735226134576y,\\
&&h_3=-104xy+335y-335x+208.
\end{eqnarray*}
\end{Example}

\section{Acknowledgments}
The authors acknowledge the support provided by NSFC-60573007,
NSFC-90718041, NKBRPC-2004CB318003 and NKBRPC-2005CB321902.


\begin{thebibliography}{99}

\bibitem{ABS94} A. G. Akritas, A. V. Bocharov and A. W.
Strzebo\'{n}ski: Implementation of real root isolation algorithms in
Mathematica. In: {\it Abstracts of the International Conference on
Interval and Computer-Algebraic Methods in Science and Engineering
(Interval¡¯94)}, 23--27. St. Petersburg, Russia, March 7-10, 1994.



\bibitem{CGY07}J. S. Cheng, X. S. Gao and C. K. Yap:
Complete Numerical Isolation of Real Zeros in General Triangular
Systems. In: {\it Proc. ISSAC2007}, 92--99, 2007.

\bibitem{co75}
  G. E. Collins: Quantifier elimination for real closed fields
  by cylindrical algebraic decomposition. In: \emph{Automata Theory and
  Formal Languages} (Brakhage, H., ed.), LNCS {\bf 33}, 134--165.
  Springer, Berlin Heidelberg, 1975.

\bibitem{CA76} G. E. Collins and A.G. Akritas: {Polynomial real root isolation
using Descartes¡¯ rule of signs.} In: {\it Proceedings of the 1976
ACM Symposium on Symbolic and Algebraic Computations}, 272--275.
Yorktown Heights, N.Y., 1976.


\bibitem{cl83} G.\,E. Collins and R. Loos:
  Real zeros of polynomials. In:
  \emph{Computer Algebra: Symbolic and Algebraic Computation}
  (Buchberger, B., Collins, G.\,E., Loos, R., eds.), 83--94.
  Springer, Wien New York, 1982.

\bibitem{cox}D. Cox, J. Little and D. O'Shea: {\it Using Algebraic
Geometry}. New York: Springer, 1998.

\bibitem{zhonggang} B. H. Dayton and Z. G. Zeng: Computing the
Multiplicity Structure in Solving Polynomial Systems. In {\it
Proceedings of the ACM-SIGSAM 2005 International Symposium on
Symbolic and Algebraic Computation}, ACM Press, 116--123, 2005.


\bibitem{gcl92} K. O. Geddes, S. R. Czapor and
G. Labahn: \emph{Algorithms for Computer Algebra}. Kluwer, Boston,
1992.

\bibitem{gg}G.-M. Greuel and G. Pfister: {\it A Singular introduction to
commutative algebra}. Springer-Verlag, Berlin, 2002.

\bibitem{Johnson} J. R. Johnson: Algorithms for Polynomial Real
Root Isolation. In: {\it Quantifier Elimination and Cylinderical
Algebraic Decomposition} (B. F. Caviness and J. R. Johnson eds.),
Springer-Verlag, 269--299, 1998.




\bibitem{RZ04} F. Rouillier and P. Zimmermann: Efficient isolation of
polynomial's real roots. {\it J. of Computational and Applied
Mathematics} \textbf{162}: 33--50, 2004.

\bibitem{sha} V. Sharma: Complexity of real root isolation using
continued fractions. In: {\it Proc. ISSAC'07}, 339--346, 2007.


\bibitem{vg99} J. von zur Gathen and J. Gerhard:
\emph{Modern Computer Algebra}. Cambridge University Press,
Cambridge, 1999.

\bibitem{wang} D. Wang: Elimination methods. Springer, Wien New
York, 2001.


\bibitem{xia07} B. Xia: DISCOVERER: A tool for solving semi-algebraic
systems. Software Demo at ISSAC 2007, Waterloo, 2007. Also: {\it ACM
SIGSAM Bulletin}, {\bf 41}(3), 102--103, 2007.

\bibitem{xh02}
B. Xia and X. R. Hou: A Complete Algorithm for Counting Real
Solutions of Polynomial Systems of Equations and Inequalities. {\it
Computers and Mathematics with Applications}, {\bf 44}: 633--642,
2002.

\bibitem{xy02}
  B. Xia and L. Yang: An algorithm for isolating the real solutions
  of semi-algebraic systems. {\it J. Symb. Comput.}, {\bf 34}:
  461--477, 2002.

\bibitem{xz06} B. Xia and T. Zhang: Real Solution Isolation Using Interval
Arithmetic. {\it Computers and Mathematics with Applications}, {\bf
52}: 853--860, 2006.


\end{thebibliography}
\end{document}